\documentclass[12pt, letterpaper]{article}
\usepackage[utf8]{inputenc}
\usepackage[margin=1in]{geometry}

\usepackage{amsmath}
\usepackage{graphicx}
\usepackage{textgreek}
\usepackage{authblk}
\usepackage[T1]{fontenc}
\usepackage{cite}

\author[1*]{Marcela Niemczyk}
\author{D. Robert Iskander}
\affil[1]{Department of Biomedical Engineering, Wroclaw University of Science and Technology, Wybrzeze~Wyspianskiego 27, 50-370 Wroclaw, Poland}
\affil[*]{\normalfont{\fontfamily{qcr}\selectfont
		{marcela.niemczyk@pwr.edu.pl}}}

\title{\textbf{Statistical modeling of corneal OCT speckle. A~distributional model-free approach}}
\date{}

\begin{document}

\maketitle

\noindent
\textbf{Abstract} \vspace{3pt}

\noindent
In biomedical optics, it is often of interest to statistically model the amplitude of the speckle using some distributional models with their parameters acting as biomarkers. In~this paper, a paradigm shift is being advocated in which a distributional model-free approach is used. Specifically, a range of distances, evaluated in different domains, between an empirical nonparametric distribution of the normalized speckle amplitude sample and the benchmark Rayleigh distribution, is considered. Using OCT images from phantoms, two \textit{ex-vivo} experiments with porcine corneas and an \textit{in-vivo} experiment with human corneas, an evidence is provided  that the distributional model-free approach, despite its simplicity, could lead to better results than the best-fitted (among a range of considered models) distributional model.  Concluding, in practice, the distributional model-free approach should be considered as the first choice to speckle modeling before a distributional-based approach is utilized.

%%%%%%%%%%%%%%%%%%%%%%%%%%  body  %%%%%%%%%%%%%%%%%%%%%%%%%%
\section{Introduction}
Speckle in optical coherence tomography (OCT) is usually treated as a source of intrinsic interference that needs to be reduced or eliminated \cite{Schmitt1999,Zhao2018}. Recently, a substantial interest in treating the OCT speckle as the source of information has been observed \cite{Kirillin2014,Almasian2017,Jesus2017a,Iskander2020,Niemczyk2021}. In such an approach, a particular local part of an OCT pseudo B-scan corresponding to the speckle amplitude signal, often termed a region of interest or a region of analysis, is treated as a random variable when the first-order statistics are considered. There is a convincing theoretical evidence, when employing the central limit theorem, that when the number of scattering elements in an~observed sample is sufficiently large, such a random variable should be Rayleigh distributed, whereas when that number of scatters is small, its statistics could be well approximated by the K~distribution \cite{Almasian2017}. Despite that, a considerable effort has been made to model the OCT speckle amplitude with other types of distributions, often of-the-shelf and physically less justified. Examples of such, in an alphabetical order, include: Bessel K form \cite{Samieinasab2020}, Burr \cite{Ge2021}, gamma \cite{Lindenmaier2013}, generalized extreme value \cite{Shetty2017}, generalized gamma \cite{Jesus2017a,Danielewska2021}, log-normal \cite{Duncan2008}, Nakagami \cite{Jesus2017a}, Rician \cite{Cheng2014}, and Weibull \cite{Shetty2017,Kajic2011} distributions. The main justification for using those models are their goodness-of-fit properties, which, in particular application, may achieve superior fitting performance to that of the two theoretically justified models \cite{Jesus2017a,Ge2021,Danielewska2021}, i.e., the Rayleigh and K~distributions.

The purpose of modeling statistical properties of OCT speckle is to characterize its random amplitude with a set of distributional parameters that might be used for a particular practical task such as, for example, discriminating between different tissue samples. In that sense, in biomedical optics, the distributional parameters could then be viewed as biomarkers. However, when faced with the choice of an optimal (in some sense) distributional model, focus seems to be mainly given to the goodness-of-fit properties of a given model rather than to its physical correctness or the statistical power of model parameters that should subsequently act as biomarkers. Such an approach favors distributions with a larger number of parameters than one, even in cases when information criteria, such as the Akaike Information Criterion \cite{Jesus2017a,Ge2021}, are employed to account for over-parameterization. Further, in order to fit the amplitude of the OCT speckle with a particular distributional model, one needs to use a~parameter estimation procedure, often based on the method of maximum likelihood, leading in case of distributions with more than one parameter to correlated parameter estimates, because orthogonality of the distributional parameters is rarely maintained. In~fact, constructing distributions with more than two orthogonal parameters is not practical because the complexity of the differential equations involved in such a construction is high \cite{Huzurbazar1950}. Hence, fitting of a three-parameter distribution, such as the Bessel K form \cite{Samieinasab2020} or the generalized gamma distribution \cite{Jesus2017a,Danielewska2021}, to the random OCT speckle amplitude can lead to a set of non-unique solutions, because different sets of parameter estimates could result in similarly well-fitted distributional representations. However, it should be emphasized that a~better goodness-of-fit achieved with a particular distribution does not necessarily correspond to a~better discriminating power of its parameter estimators \cite{Iskander2020}.
 
In this paper, a paradigm shift is advocated to deviate from the parametric distributional approach to modeling the amplitude of the OCT speckle. Instead, after speckle amplitude normalization, the approach based on the distance between the Rayleigh distribution with a fixed parameter $\sigma$ =$\sqrt{2}/2$ and an empirical nonparametric distribution of the sample is considered. Such distances are being evaluated in the domain of probability density function, cumulative distribution function as well as that of the characteristic function. Also, the difference (distance) between the sample contrast ratio and the theoretical contrast ratio value for the Rayleigh distribution, is considered. To support our developments, OCT images from purposely designed phantoms, \textit{ex-vivo} examination of porcine corneas and \textit{in-vivo} examination of human corneas, are utilized. Further, the inefficiency of the distributional-based approach to speckle modeling is exposed for the case of a distributional model achieving, among other considered distributions, the best goodness-of-fit.

\section{Methods}

Before introducing the incorporated image analysis methods involved in the proposed distributional model-free approach to speckle modeling, a summary of speckle theory and the associated statistical tools is first provided, for completeness.

\subsection{Preliminaries}
\label{sec:Preliminaries}
The fundamental theory of a speckle pattern formation was described by Goodman \cite{Goodman1975,Goodman2007}. It is assumed that the electric field amplitude at some observation point of a speckle pattern results from contributions from different regions of the scattering media. The resultant complex amplitude is considered as a superposition of the elementary phasors with their amplitudes and phases statistically independent from each other and from amplitudes and phases of other phasors. The phases are also expected to be uniformly distributed $U(-\pi,\pi)$. Having these assumptions, the real and imaginary parts of the resultant amplitude are considered to have zero mean, equal variances, and be uncorrelated. So if the number of elementary phasor contributions is large, the real and imaginary parts of the resultant phasor follow the Gaussian distribution, in compliance with the central limit theorem. It can be proven, that the length of the two-dimensional vector, which components are normally distributed with zero means and equal variances, $\sigma ^2$, obeys Rayleigh statistics. Therefore, the amplitude (length) $A$ of the resultant phasor in some point of the speckle field follows the Rayleigh distribution with the scale parameter $\sigma$, described by the probability density function (PDF) given by
\begin{equation}
	p_A (A) = \frac{A}{\sigma ^2}e^{-\frac{A^2}{2\sigma ^2}}\:,\quad A\geq0.
\end{equation}
The aforementioned result is approximate for a finite large number of elementary phasors and becomes valid when the number of elementary phasors tends to infinity. The speckle field is then referred to as fully developed \cite{George1976}. Additionally, for such a speckle field, the contrast ratio, defined as the ratio of the standard deviation of the amplitude to its mean value, approaches its Rayleigh-limited value of $\sqrt{4/\pi-1}\approx 0.5227$ \cite{Hillman2006}.

The calculations of speckle statistics in OCT images are usually based on pixel values that represent the speckle field amplitude $A$ in a particular point (dependent on the image resolution) \cite{Fercher2003}. Since the speckle field amplitude follows the Rayleigh distribution, it is of interest to estimate the parameter of this distribution for the pixel values. In practice, it is usually considered to normalize the amplitude by dividing it by its root mean square (RMS) value \cite{Ge2021,Stanton2018}, i.e., $A/\sqrt{\langle A^2 \rangle}$. Such normalization, in many cases, leads to simplified parameter estimation procedures, in which one of the parameters (i.e., that of scale) achieves a particular constant value. It is worth noting that estimating the scale parameter for the normalized speckle amplitude may lead to incorrect interpretation of the estimated value because that estimator is biased and that bias depends on the other distributional parameter estimates (i.e., those of shape). 

There are several methods available for estimating the scale parameter of the Rayleigh distribution. Because of its properties, the maximum likelihood estimator (MLE) is most frequently used and it takes the form
\begin{equation}
	\hat{\sigma}_{\mathrm{MLE}}=\sqrt{\frac{1}{2n}\sum_{i=1}^{n}A_i^2}\:,
\end{equation}
where $A_i=1,2,\dots, n$, with $n$ being the number of samples, are the discrete samples of the speckle amplitude random variable $A$. If the amplitude of the speckle pattern is normalized, the above-mentioned estimator reduces to
\begin{equation}
	\hat{\sigma}_{\mathrm{MLE}}=\sqrt{\frac{1}{2n}\cdot \frac{1}{\frac{1}{n}\sum_{i=1}^{n}A_i^2}\cdot \sum_{i=1}^{n}A_i^2}=\frac{\sqrt{2}}{2}\:,
\end{equation}
leading to the Rayleigh distribution with a fixed scale parameter. One may consider another estimator for $\sigma$ than MLE such as that proposed by Ardianti who used a Bayes method of the form \cite{Ardianti2018}
\begin{equation}
	\hat{\sigma}_{\mathrm{Bayes}} = \frac{\sqrt{2}\Gamma(n+2)}{2\Gamma\big(n+\frac{5}{2}\big)}\sqrt{\sum_{i=1}^{n}A_i^2}\:.
\end{equation}
Again, for the normalized speckle amplitude one obtains
\begin{equation}
	\hat{\sigma}_{\mathrm{Bayes}} = \frac{\sqrt{2}\Gamma(n+2)}{2\Gamma\big(n+\frac{5}{2}\big)}\sqrt{\frac{1}{\frac{1}{n}\sum_{i=1}^{n}A_i^2}\cdot \sum_{i=1}^{n}A_i^2}=\frac{\sqrt{2}}{2}\cdot \frac{\Gamma(n+2)}{\Gamma\big(n+\frac{5}{2}\big)}\cdot \sqrt{n} \:,
\end{equation}
which asymptotically, for $n\to \infty$, approaches $\sqrt{2}/2$. This indicates that for a fully developed speckle, the normalized speckle amplitude follows the Rayleigh distribution with a fixed scale parameter, regardless of the estimation method.

The distribution described above is theoretically justified for the speckle amplitude provided that the number of scatterers in an examined medium is large. However, sometimes the number of scatterers may be insufficient to obtain a fully developed speckle field. The speckle statistics are then altered, as the central limit theorem is no longer applicable. The~theoretical derivations on that problem, developed by Jakeman and Pusey \cite{Jakeman1978}, included the assumption that the number of contributions to the scattered field is fluctuating and could be modeled by a negative binomial distribution. That leads to the K distribution as a model of the speckle field amplitude with the form \cite{Jakeman1984}
\begin{equation}
p_A(A)=\frac{4}{\Gamma (\alpha)}\sqrt{\frac{\alpha}{\langle  A^2\rangle}}\cdot \Bigg( \frac{\alpha A^2}{\langle  A^2\rangle} \Bigg)^\frac{\alpha}{2}K_{\alpha-1}\Bigg(2\sqrt{\frac{\alpha A^2}{\langle  A^2\rangle}} \Bigg) \:,
\end{equation}
where $\alpha$ is the shape parameter of the K distribution. This model is widely used in works where a~small number of scatterers is considered. Also, the shape parameter of the K~distribution is linked to the average scatterer density in the tested object \cite{Popov1993,Sugita2016,Weatherbee2016}. 
Owing to the fact that the Rayleigh distribution is theoretically proven to be suitable for modeling the fully developed speckle field amplitude, in section \ref{section:OurMethod} we advocate a distributional model-free approach for speckle statistics analysis, which does not involve any distributional parameter estimation. However, first, the experimental data, used to illustrate the proposed approach and its potential applications, is described.

\subsection{Experimental data}
To validate the distributional model-free approach, data from three studies performed on resin phantoms, \textit{ex-vivo} on porcine eyeballs, and \textit{in-vivo} on human eyes are used. For each of them, the OCT B-scans were acquired using a spectral OCT (SOCT Copernicus REVO, Optopol, Zawiercie, Poland), with the center wavelength of 830~nm, the half bandwidth of 50~nm, the axial resolution of 5~\textmu m, and the transversal resolution of 15~\textmu m. The scanning speed of the device is 80 000 A-scans per second. The measurements were acquired at a~constant aperture within the bands of the instrument’s depth of focus using the own guiding system of the instrument.

\subsubsection{Phantoms}
The first study, performed on phantoms, was a pilot study to assess the impact of scatterer density on the speckle statistics and the results of the distributional model-free approach. For this, a set of purposely designed phantoms were fabricated. They were made of the epoxy resin L-285 (Havel composites, Cieszyn, Poland), which in its liquid state was thoroughly mixed with the blue dye powder particles of approximate size of a few to tens micrometers. When the dye powder was visually uniformly distributed in the resin, the mixture was carefully poured on a~microscope slide, to form a shape of a convex disc with the thickness of about 1~mm and the diameter of about 10~mm. After drying, the phantoms became solid transparent discs with visible blue particles. This procedure was repeated 9 times, increasing the amount of the powder dye, to obtain phantoms with different scattering particle concentrations (C\textsubscript{1}, C\textsubscript{2}, ..., C\textsubscript{9}). The discs’ dimensions and the concentration of the dye were not precisely controlled but the increasing trend of concentration between subsequent phantoms was well maintained. For each phantom, a~single OCT B-scan of size $3077\times 708$~pixels, encompassing the central part of the phantom, was registered.

\subsubsection{Porcine corneas --- \textit{ex-vivo} study}
Data from a recent study \cite{Niemczyk2021} consisting of two \textit{ex-vivo} experiments on porcine eyeballs, in which the parametric distributional-based approach to speckle modeling was used, were taken for validating the distributional model-free approach. Experiment~1 was performed to evaluate the influence of intraocular pressure (IOP) elevation on the corneal OCT speckle statistics. The IOP in the anterior chamber of each eyeball was increased from 10~mmHg to 40~mmHg with the step of 5~mmHg and the OCT scans of size $1536\times 736$ pixels were registered at each IOP level. Experiment~2 was prepared according to the same procedure, but with IOP maintained at the constant level of 15~mmHg. The OCT scans were collected in 7 time points, adequate to the first experiment, to examine the impact of the experiment’s duration on speckle statistics. Thirty three eyeballs were included in the study: 23 eyeballs for Experiment~1 and 10 eyeballs for Experiment~2. The entire procedure of the experiments’ preparation and the characteristics of the equipment are described in detail in that recent work \cite{Niemczyk2021}.

\subsubsection{Human corneas --- \textit{in-vivo} study}
The final set of data, used here for the validation of the distributional model-free approach to speckle modeling, were measurements of the central cornea in a group of 56 healthy Caucasian subjects with the mean ($\pm$ standard deviation) age of $42.4 \pm 18.3$ years (range from 21 to 87 years). For each subject, a single OCT B-scan of size $1536\times 736$ was acquired for randomly chosen eye and a non-corrected IOP was measured using the noncontact tonometer Corvis Scheimpflug Technology (Corvis ST, OCULUS, Wetzlar, Germany). All registered values of IOP were within the normal physiological limits (i.e., less than or equal to 20~mmHg). This part of the study was performed in accordance with the tenets of the Declaration of Helsinki.

\subsection{Image analysis}
\begin{figure}[htbp]
	\centering\includegraphics[width=12cm]{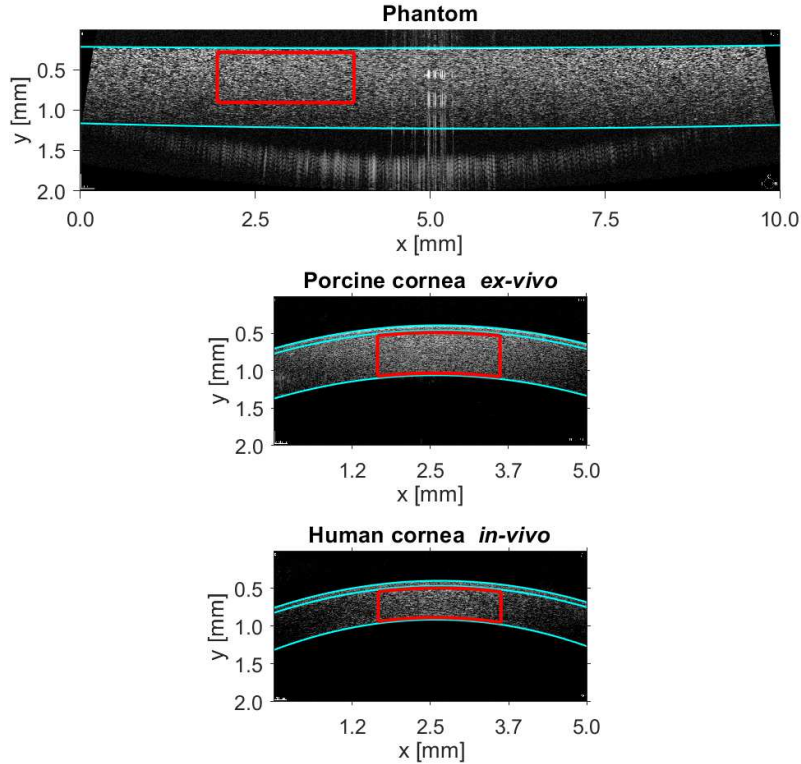}
	\caption{Illustrative OCT scans from each considered study with ROI marked with red lines. Cyan lines indicate phantom borders (top image) or epithelium, Bowman’s layer, and endothelium (the middle and bottom image).}
	\label{Fig1}
\end{figure}

All calculations were performed in MATLAB (MathWorks, Inc. Natick, MA, USA). At first, the log-transformation, applied automatically to every OCT B-scan in the device software, was inversed. Pixel values were selected from the specified region of interest (ROI), dependent on the type of imaged object (Figure \ref{Fig1}). For phantoms, the ROI was placed on the left of the central reflection, 20~pixels below the top border of the phantom, and was of size $600\times220$ pixels, encompassing an area of about 2~mm horizontally by 0.6~mm vertically. Note that for imaging phantoms, an instrument protocol with an external adapting lens was used, allowing the so-called wide scan. For porcine and human corneas the ROI was placed centrally in relation to the apex, 10~pixels below the Bowman’s layer and 10~pixels above the endothelium, and had the width of 600~pixels, which corresponds to the central 2~mm of the cornea. For pixel values from each ROI, treated as a random variable $X$, empirical cumulative distribution function (eCDF), kernel density estimator (KDE), empirical characteristic function (eCF), and the sample contrast ratio (CR), were calculated in accordance with the formulas
\begin{itemize}
	\item eCDF \cite{Vaart1998}
	\begin{equation}
		\mathrm{eCDF}=\frac{1}{n}\sum_{i=1}^{n}\textbf{1}\{X_i\leq t\} \:,
	\end{equation}
	where $\textbf{1}\{X_i\leq t\}$ is the indicator of the event that the value of a random variable $X_i$ is less than or equal to $t$, whereas $n$ is the number of samples in ROI.
	
	\item KDE \cite{Silverman2018}
	\begin{equation}
	\mathrm{KDE}=\frac{1}{n}\sum_{i=1}^{n}\frac{1}{h}K\Big(\frac{x-X_i}{h}\Big)\:,
	\end{equation}
	where $h$ is the bandwidth of the estimator and $K$ is the non-negative kernel function. The Gaussian type of kernel function was set and the bandwidth value was $h=(0.75n)^{(-1/5)}\hat{\sigma}_X$ with $\hat{\sigma}_X$ being the sample standard deviation. In KDE calculations it was necessary to correct a boundary effect, related to the left side of the interval of pixel values, which are nonnegative. For this purpose, KDE values for abscissa less than 0.05 were not taken into account in further calculations \cite{Jones1993}.
	
	\item eCF \cite{Cramer1946}
	\begin{equation}
	\mathrm{eCF}=\frac{1}{n}\sum_{i=1}^{n}e^{jtX_i}\,, \quad  j=\sqrt{-1}\:.
	\end{equation}
	
	\item CR \cite{Goodman2007}
	\begin{equation}
	\mathrm{CR} = \frac{\hat{\sigma}_X}{\overline{X}}\:,
	\end{equation}
	where $\overline{X}$ denotes the sample mean. 	
\end{itemize}

\subsection{A distributional model-free approach}
\label{section:OurMethod}
In this work, a distributional model-free approach for OCT corneal speckle analysis is advocated. Taking into account that the theoretical speckle amplitude follows the Rayleigh distribution and that for the normalized amplitude its parameter has a constant value, the Rayleigh distribution with a scale parameter $\sigma=\sqrt{2}/2$ is used as a benchmark distribution (BD). The empirical distributions of pixel values from the specified ROI in OCT images are compared with this model using the following four different statistical distances:
\begin{itemize}
	\item Kolmogorov–Smirnov distance between eCDF and the cumulative distribution function (CDF) of the benchmark distribution \cite{Massey1951} 
	\begin{equation}
		D_{\mathrm{KS}}=\sup_x|\mathrm{eCDF}(x)-\mathrm{CDF_{BD}}(x)|
	\end{equation}
	where $\sup_x$ is the supremum of the set of distances across all $x$ values.

	\item mean square error (MSE) distance, between KDE and the probability density function (PDF) of the benchmark distribution
	\begin{equation}
		D_{\mathrm{MSE}} = \frac{1}{n}\sum_{i=1}^{n}(\mathrm{KDE}(x_i)-\mathrm{PDF_{BD}}(x_i))^2
	\end{equation}
	
	\item maximum mean discrepancy (MMD) distance, between eCF and the characteristic function (CF) of the benchmark distribution \cite{Gretton2012}
	\newcommand{\norm}[1]{\left\lVert#1\right\rVert}
	\begin{equation}
		D_{\mathrm{MMD}} \norm{\frac{1}{n}\sum_{i=1}^{n}\mathrm{eCF}(x_i)-\frac{1}{n}\sum_{i=1}^{n}\mathrm{CF_{BD}}(x_i)}
	\end{equation}
	
	\item CR distance, between contrast ratio calculated from pixel values and the theoretical value of 0.5227 for the Rayleigh distribution 
	\begin{equation}
		D_{\mathrm{CR}}=\mathrm{CR}-0.5227
	\end{equation}
\end{itemize}

\subsection{Statistical analysis}
In the \textit{ex-vivo} study on porcine corneas, changes in statistical distance values were evaluated for two experiments. One way repeated measures analysis of variance (rmANOVA) was used to assess whether the values of considered statistical distances vary with IOP (Experiment~1) and with time (Experiment~2). Post-hoc analysis was involved to assess the statistical significance of changes in distance values between adjacent IOP levels or time points. To obtain the correlation coefficient between the statistical distance and IOP or time, linear mixed effect model fitting was applied, because for the same eyeball distance values for consecutive IOP levels or time points are correlated. In \textit{in-vivo} study on human corneas, Pearson correlation coefficients were calculated between distance values and IOP. Differences between correlation coefficients were tested using the Fisher test.

\section{Results}
\subsection{Phantoms}
\begin{figure}[htbp]
	\centering\includegraphics[width=11cm]{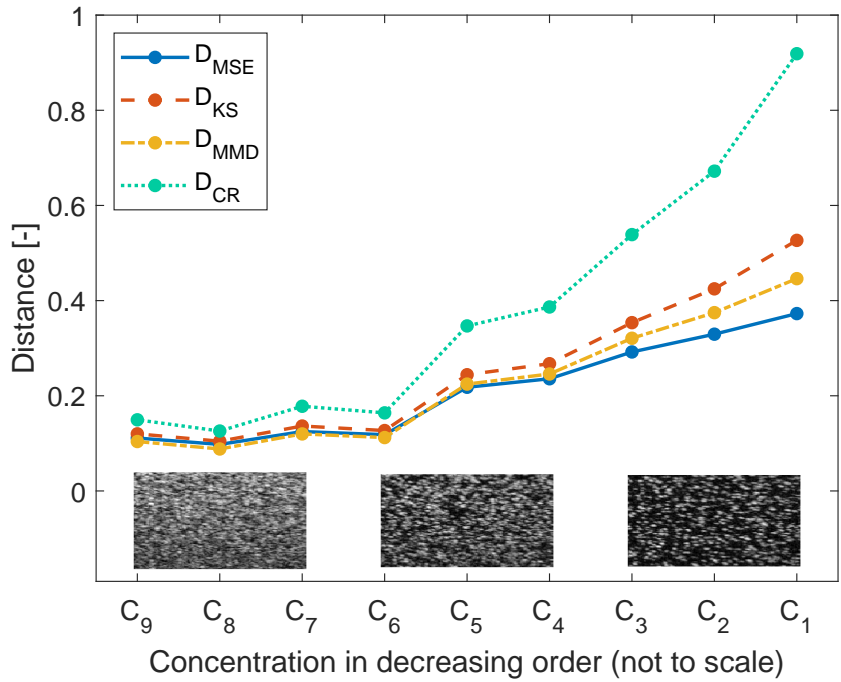}
	\caption{Distance values for phantoms with decreasing concentration of blue dye particles. Illustrative images inserted at the bottom of the plot present the ROIs of the OCT images of the phantoms for the concentrations C\textsubscript{8}, C\textsubscript{5}, and C\textsubscript{2}.}
	\label{Fig2}
\end{figure}

\begin{figure}[htbp]
	\centering\includegraphics[width=11cm]{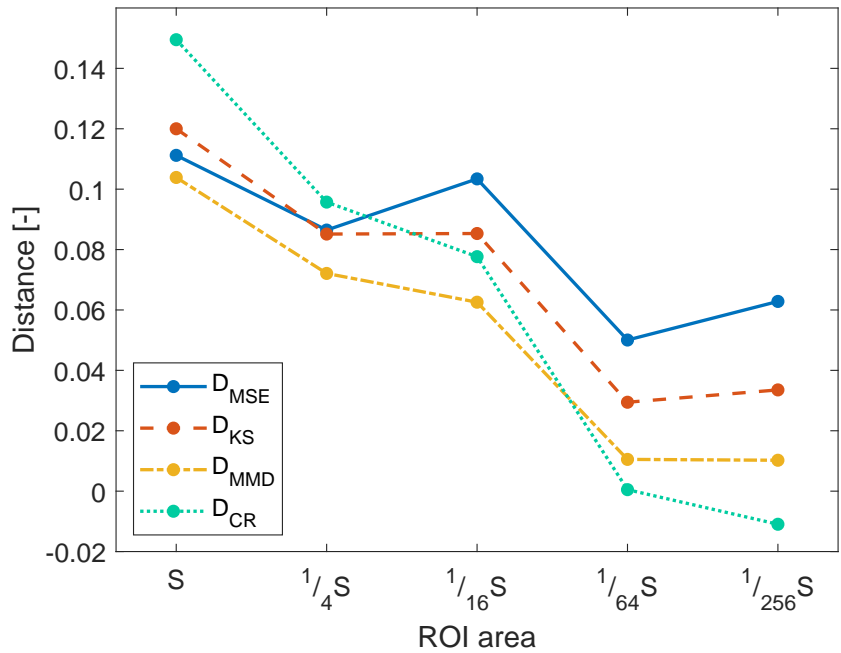}
	\caption{Estimated distances as functions of ROI area for the phantom of concentration~C\textsubscript{9}. S~denotes the original ROI area of $600\times220$ pixels.}
	\label{Fig3}
\end{figure}

Measurements of the fabricated phantoms were used to assess whether the scatterer density has an impact on the statistical distance values. Figure \ref{Fig2} shows that the decrease in scatterer concentration causes deterioration of the Rayleigh model fitting to experimental data, meaning that the empirical distribution becomes more divergent from the benchmark distribution. Hence, the Rayleigh distribution with a scale parameter $\sqrt{2}/2$ becomes more valid when the number of scatterers is larger, which is consistent with theoretical derivations on speckle field amplitude, described in section \ref{sec:Preliminaries}. Contrast ratio distance, as well as other distances, have similar, less than 0.2, values for higher dye concentrations (C\textsubscript{9} -- C\textsubscript{6}) and starts to increase from concentration C\textsubscript{5} achieving highest values for concentration C\textsubscript{1}. It is worth noting that by reducing the size of ROI to a small local neighborhood \cite{Duncan2008}, for a~fully developed speckle one can achieve distance values approaching zero. Figure \ref{Fig3} shows the estimated distances as functions of ROI size for the phantom of concentration C\textsubscript{9}.

\subsection{Porcine corneas --- \textit{ex-vivo} study}
\begin{figure}[htbp]
	\centering\includegraphics[width=15cm]{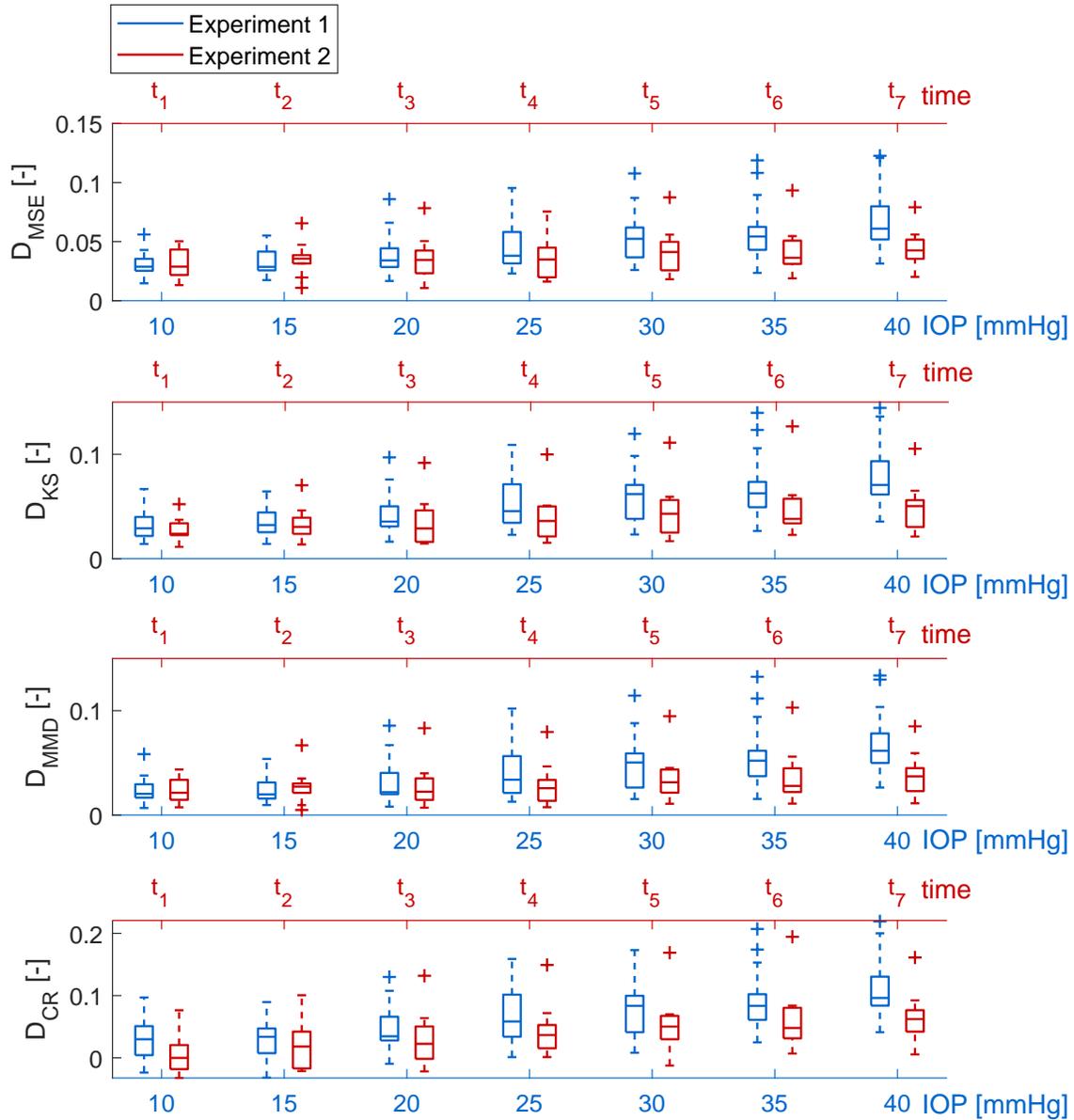}
	\caption{Boxplots of $D_\mathrm{MSE}$, $D_\mathrm{KS}$, $D_\mathrm{MMD}$, and $D_\mathrm{CR}$ values for Experiment~1 (examining the impact of IOP) and Experiment~2 (examining the impact of its duration) on porcine corneas (\textit{ex-vivo} study).}
	\label{Fig4}
\end{figure}

Figure \ref{Fig4} presents the boxplots of the statistical distance values for two \textit{ex-vivo} experiments on porcine corneas. Repeated measures ANOVA showed that there are statistically significant differences between distance values for different IOP levels ($p < 0.001$ for all considered distances) and for different time points ($p = 0.024$, $p < 0.001$, $p = 0.004$, and $p < 0.001$ for $D_\mathrm{MSE}$, $D_\mathrm{KS}$, $D_\mathrm{MMD}$, and $D_\mathrm{CR}$, respectively). Post-hoc analysis was performed to assess statistical significance of distance values differences between adjacent IOP levels and consecutive time points. It revealed that in Experiment~2 there are statistically significant differences in three of considered distances values between $t_4$ and $t_5$ time points ($p = 0.024$, $p = 0.050$ and $p = 0.020$ for $D_\mathrm{MSE}$, $D_\mathrm{KS}$ and $D_\mathrm{MMD}$, respectively) and in $D_\mathrm{KS}$ values between $t_3$ and $t_4$ ($p = 0.043$). Statistically significant distances are also observed in $D_\mathrm{CR}$ values for pairs of adjacent time points from $t_1$ to $t_4$ ($p < 0.001$, $p = 0.030$, $p = 0.004$). In Experiment~1, there are statistically significant differences in values of all distances between adjacent IOP levels from 15~mmHg to 35~mmHg (consecutive \textit{p}-values were: for $D_\mathrm{MSE}$ 0.049, 0.001, 0.001, 0.008; for $D_\mathrm{KS}$ 0.025, 0.0003, 0.001, 0.002; for $D_\mathrm{MMD}$ 0.023, 0.001, 0.002, 0.003; for $D_\mathrm{CR}$ $<0.001$, $<0.001$, 0.002, 0.002). The correlation coefficients between statistical distances and IOP, calculated using linear mixed-effect model fitting, are 0.841, 0.864, 0.844 and 0.880 for $D_\mathrm{MSE}$, $D_\mathrm{KS}$ $D_\mathrm{MMD}$ and $D_\mathrm{CR}$, respectively.

\subsection{Human corneas --- \textit{in-vivo} study}
\begin{figure}[htbp]
	\centering\includegraphics[width=12cm]{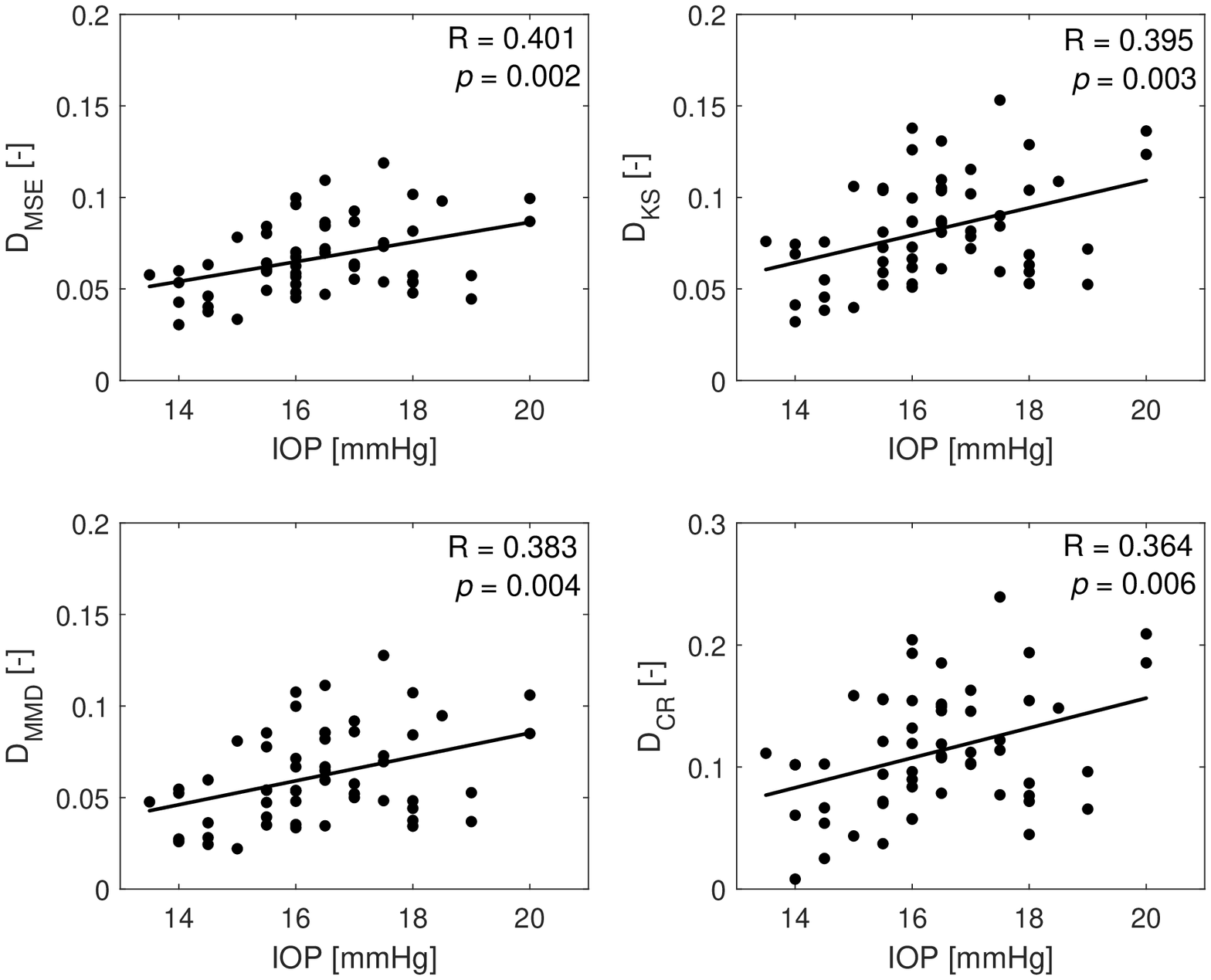}
	\caption{$D_\mathrm{MSE}$, $D_\mathrm{KS}$, $D_\mathrm{MMD}$ and $D_\mathrm{CR}$ values for the OCT images of human corneas (\textit{in-vivo} study). The lines present linear regression and the R values are the correlation coefficients between IOP and distance values.}
	\label{Fig5}
\end{figure}

The \textit{in-vivo} study on human corneas showed that there are statistically significant correlations between IOP and values of statistical distances (Figure \ref{Fig5}). The values of Pearson correlation coefficient were 0.401, 0.395, 0.383, 0.364 and corresponding \textit{p}-values were 0.002, 0.003, 0.004, and 0.006 for $D_\mathrm{MSE}$, $D_\mathrm{KS}$, $D_\mathrm{MMD}$, and $D_\mathrm{CR}$, respectively. The correlations were weak but showed that the overall trend of statistical distance values increasing with IOP, found in the porcine corneas in the \textit{ex-vivo} study, is preserved. It is worth noting that the maximum correlation coefficient value of 0.401, achieved for $D_\mathrm{MSE}$, is not statistically significantly different to the minimum correlation coefficient value of 0.364, achieved for $D_\mathrm{CR}$ (Fisher test, $p = 0.396$).

\subsection{Ineffectiveness of the distributional-based approach}
\begin{figure}[htbp]
	\centering\includegraphics[width=15cm]{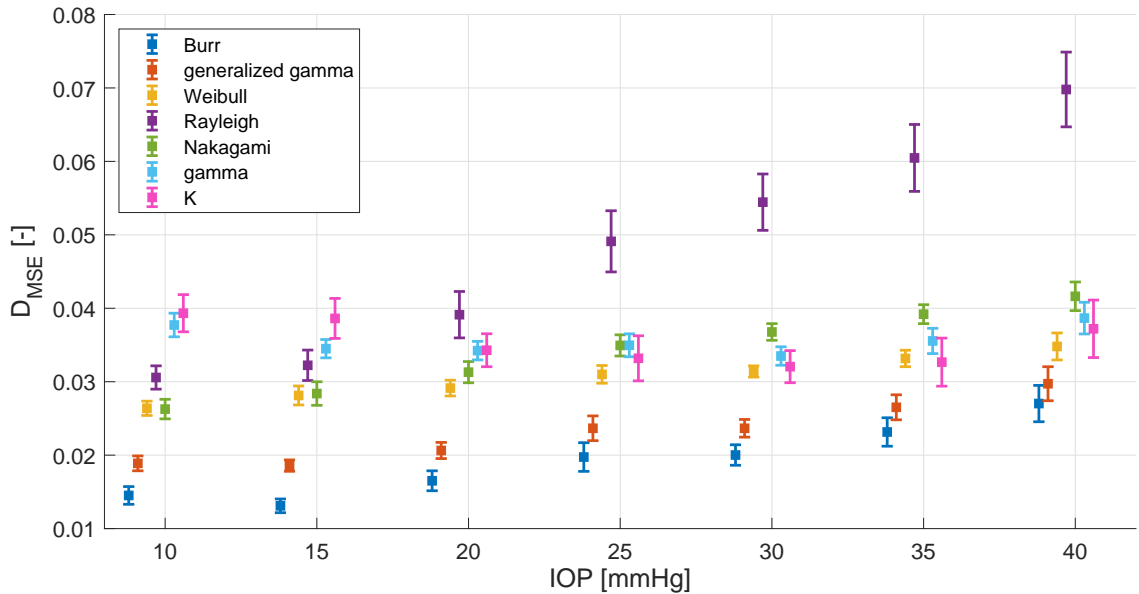}
	\caption{Means and standard errors of the goodness-of-fit (GoF) for a set of different distributions for the OCT images of porcine eyeballs in the \textit{ex-vivo} study in Experiment~1.}
	\label{Fig6}
\end{figure}

To demonstrate the ineffectiveness of the distributional-based approach, it is worth showing that a better goodness-of-fit achieved with a particular distribution does not necessarily correspond to a better discriminating power of its parameter estimators. For this,  a set of seven models, namely the Burr, gamma, generalized gamma, K, Nakagami, Rayleigh, and Weibull distributions were used to fit the data from the first \textit{ex-vivo} porcine eye study (Experiment~1). Further, for each measurement, the goodness-of-fit (GoF), here corresponding to the mean square error between the given estimated PDF and the KDE, was calculated. The results of the GoF are shown in Figure~\ref{Fig6}. It is evident that the two three-parameter distributions, namely, the generalized gamma and the Burr, achieve better GoF results than those of one- or two-parameter distributions. 

\begin{figure}[htbp]
	\centering\includegraphics[width=16cm]{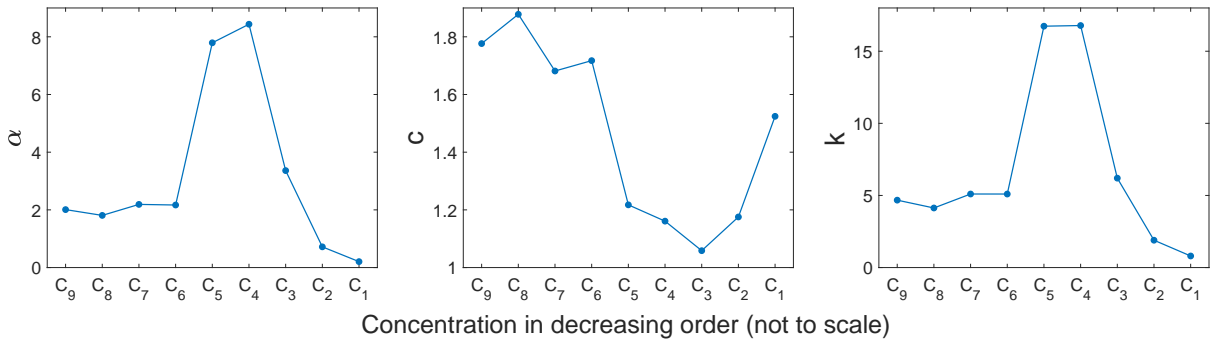}
	\caption{Values of the Burr distribution parameters for decreasing concentrations of dye particles in the phantom study.}
	\label{Fig7}
\end{figure}

\begin{figure}[htbp]
	\centering\includegraphics[width=13cm]{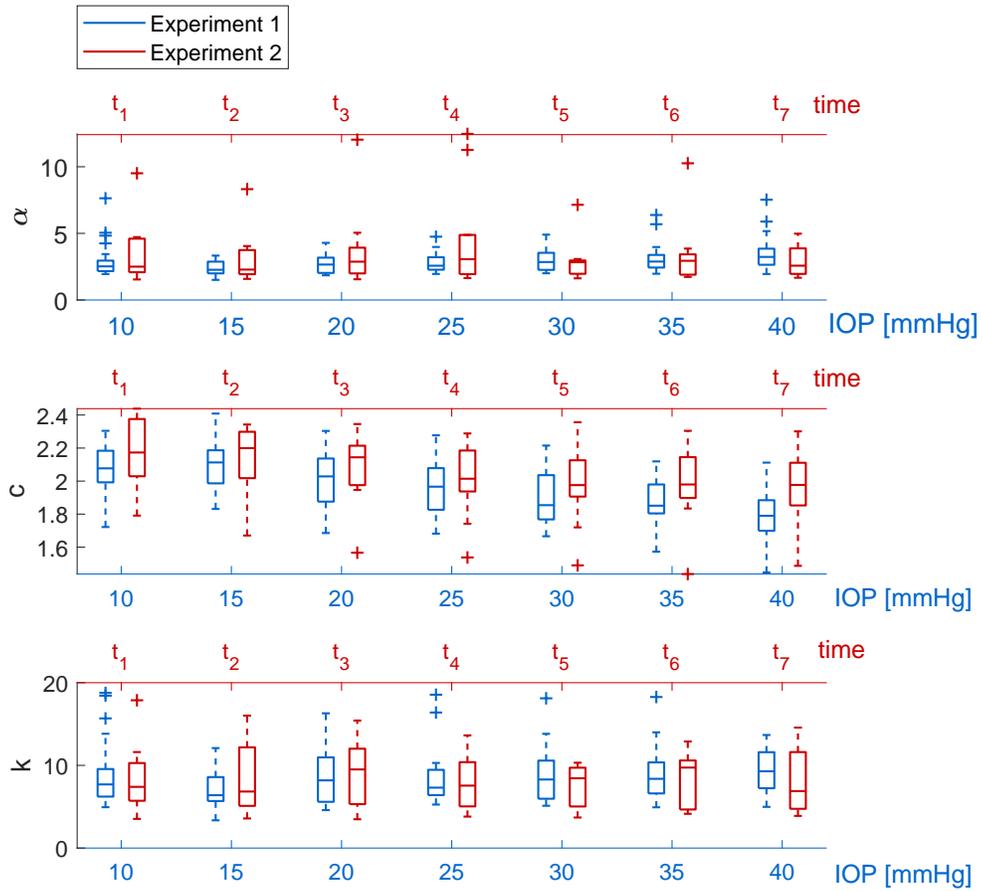}
	\caption{Boxplots of the Burr distribution parameters for Experiment~1 (examining the impact of IOP) and Experiment~2 (examining the impact of its duration) on porcine corneas (\textit{ex-vivo} study).}
	\label{Fig8}
\end{figure}

\begin{figure}[htbp]
	\centering\includegraphics[width=17cm]{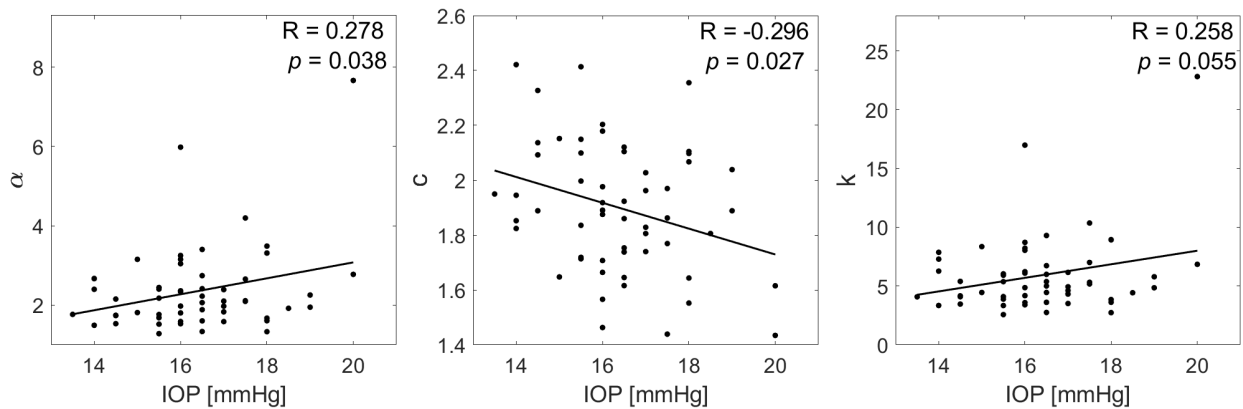}
	\caption{Values of the Burr distribution parameters for the OCT images of human corneas (\textit{in-vivo} study). The lines present linear regression and the R values are the correlation coefficients between IOP and parameters.}
	\label{Fig9}
\end{figure}

For further analysis, the Burr distribution is chosen because it achieved the best GoF result among the distributions that were considered. The PDF of three-parameter Burr distribution is
\begin{equation}
	p_A(A)=\frac{kc}{\alpha}\Big(\frac{A}{\alpha} \Big)^{c-1} \Big(1+\Big(\frac{A}{\alpha} \Big)^c \Big)^{-k-1}\:,
\end{equation}
where $\alpha>0$ is the scale parameter, whereas $c>0$ and $k>0$ are the two shape parameters. The distributional parameters were estimated using the method of maximum likelihood. Figures \ref{Fig7}, \ref{Fig8}, and \ref{Fig9} show the behavior of the parameters of the Burr distribution for the three considered sets of data from phantoms, \textit{ex-vivo} porcine eye study, and \textit{in-vivo} human corneal study, respectively. It is evident, as noted earlier, that the distributional-based approach, even for the best case of GoF, when compared to the advocated here distributional model-free approach (see Figures \ref{Fig2}, \ref{Fig4}, and~\ref{Fig5}), is ineffective for the considered studies. In~particular, for phantoms, none of the estimated distributional parameters (i.e., $\alpha$, $c$ or $k$) show any useful trend with the sample concentration whereas for the porcine corneas \textit{ex-vivo} study, no statistically significant differences between different levels of IOP or different time instances were found. Also, for the human corneas \textit{in-vivo} study, the correlations between the parameter estimates and IOP are lower than those achieved for the distributional model-free distances. Those limitations of the distributional-based approach stem from the fact that the model parameter estimates are not independent of each other. 

\section{Discussion}
In this paper, several statistical distances based on PDF, CDF, and CF, were employed to quantify the similarity of the empirical distribution of the normalized speckle amplitude to the benchmark Rayleigh distribution with the scale parameter $\sigma = \sqrt{2}/2$. Additionally, a distance between the sample contrast ratio and that of the Rayleigh distribution was considered. The Rayleigh distribution is an appropriate physically justified model for speckle field amplitude provided that the number of scatterers is relatively large. Hence, it is concluded that increasing the number of scatterers in a tested object should tend the empirical distribution of the speckle amplitude to the Rayleigh distribution, minimizing the statistical distance between those two distributions. Such an approach has been used by Matveev et al. \cite{Matveev2019}, who used a Rayleigh distribution goodness-of-fitting to isolate low scattering biological structures and to distinguish them from noise. Here, this study additionally contributes to those developments, showing that statistical distance measures, independent of the domain that are calculated in, can be successfully utilized to distinguish between different samples as well as different sample conditions. 

Regarding the contrast ratio, which value for a fully developed speckle field, that can be modeled by the Rayleigh distribution, is approximately equal to 0.52, theoretical and experimental results, described by Hillman et al. \cite{Hillman2006}, show the negligible variation in the contrast ratio for a~large number of scatterers and strong variation of its values for lower number of scatterers. The use of contrast ratio for scatterer density characterization, especially for the speckle fields that are not fully developed, was proposed earlier \cite{Duncan2008,Lamouche2012}. The results of the phantom experiment considered in this study confirm those developments. The rationale behind using a statistical distance rather than a particular distributional model itself is evident when considering the results for the fabricated phantoms. It is clear that for low concentration levels (C\textsubscript{1}, C\textsubscript{2}, ..., C\textsubscript{5}) the speckle pattern is not fully developed (the CR distance does not approach zero) whereas for higher concentration levels (C\textsubscript{6}, C\textsubscript{7}, ..., C\textsubscript{9}) that speckle approaches that of a fully developed pattern. 

In the \textit{ex-vivo} study on porcine eyes the increase of statistical distance values was observed during the IOP elevation. This suggests that the increase in IOP causes the decrease in scatterer density and that is in agreement with the study of Wu et al. \cite{Wu2008} who showed similar evolution of collagen microstructure, in particular the immediate loss of interlamellar gaps, with increasing IOP using nonlinear optical microscopy. Hence, the evaluation of statistical distances for the speckle pattern can provide some insight into the microstructure of the imaged tissue sample. Furthermore, the \textit{in-vivo} study on human corneas confirms the positive trend of statistical distance values with increasing IOP that has been shown earlier in the porcine \textit{ex-vivo} study.

Further, the distributional model-free approach has been contrasted against a distributional-based approach that achieved the best goodness-of-fit results among seven popular models of OCT speckle. This comparison clearly exposed the limitations of the distributional-based approach, whose parameters, unlike the advocated here statistical distributional model-free measures, could not be used for assessing the differences between samples considered. Despite those limitations, it has to be acknowledged that in some applications the distributional-based approach showed promising results \cite{Jesus2017a,Danielewska2021,Ge2021}. In those cases, of interest would be to contrast those results against advocated here statistical distance measures, which among discussed earlier advantages possess higher computational efficiency than those of the distributional-based techniques.  

Summarizing, when considering the first order statistics of the speckle amplitude, the distance between the sample contrast ratio and that of the Rayleigh distribution is the simplest, yet equally effective measure of speckle departure from a fully developed field than other statistical distances evaluated in different domains. This study emphasizes that such a distributional model-free approach should be considered before any other, often more complicated, distributional-based approach is utilized.	

\newpage
\noindent
\textbf{Acknowledgments} \vspace{3pt}

\noindent
Authors wish to express their gratitude and thank Dr Monika Danielewska and Dr~Małgorzata Kostyszak for their joint contributions to the three considered studies included in this work as well as for the fruitful discussions.
\noindent

%%%%%%%%%%%%%%%%%%%%%%% References %%%%%%%%%%%%%%%%%%%%%%%%%
\bibliographystyle{unsrt}
\bibliography{Bibliography}

\end{document}